\documentclass{article}

\usepackage{times}
\usepackage{amssymb,amsmath,amsthm}
\usepackage{mathrsfs}

\DeclareMathAlphabet\EuScript{U}{eus}{m}{n}
\SetMathAlphabet\EuScript{bold}{U}{eus}{b}{n}

\renewcommand{\ae}{\emph{a.e.}~}
\newcommand{\ie}{\emph{i.e.}~}
\newcommand{\eg}{\emph{e.g.}~}

\def\B{\mathscr{B}}
\def\C{\mathbb{C}}
\def\D{\mathscr D}
\def\E{\mathscr E}
\def\F{\mathscr F}
\def\G{\mathcal G}
\def\H{\mathcal H}
\def\K{\mathcal K}

\def\N{\mathbb{N}}

\def\R{\mathbb{R}}

\def\<{\left\langle}
\def\>{\right\rangle}
\def\({\left(}
\def\){\right)}
\def\[{\left[}
\def\]{\right]}
\def\ltwo{\mathsf{L}^{\:\!\!2}}
\def\lone{\mathsf{L}^{\:\!\!1}}
\def\lthreeloc{\mathsf{L}^{\:\!\!3}_{\rm loc}}
\def\linf{\mathsf{L}^{\:\!\!\infty}}
\def\linfloc{\mathsf{L}^{\:\!\!\infty}_{\rm loc}}
\def\d{\mathrm{d}}
\def\e{\mathrm{e}}
\def\essinf{\mathop{\mathrm{ess\,inf}}}
\def\esssup{\mathop{\mathrm{ess\,sup}}}
\def\dom{\matheucal D}

\def\supp{\mathop{\mathrm{supp}}\nolimits}

\newtheorem{Theorem}{Theorem}[section]
\newtheorem{Proposition}[Theorem]{Proposition}
\newtheorem{Lemma}[Theorem]{Lemma}
\newtheorem{Remark}[Theorem]{Remark}

\newtheorem{Assumption}[Theorem]{Assumption}
\newtheorem{Definition}[Theorem]{Definition}

\begin{document}


\title{{\Large\textbf{On the spectrum of magnetic Dirac operators \\ with Coulomb-type
perturbations}}}

\author{Serge Richard$^1$ and Rafael Tiedra de Aldecoa$^2$}
\date{\small
\begin{quote}
\emph{
\begin{itemize}
\item[$^1$] Universit\'e de Lyon, Lyon, F-69003, France; Universit\'e Lyon 1, Institut Camille Jordan,
Villeurbanne Cedex, F-69622, France; CNRS, UMR5208, Villeurbanne
Cedex, F-69622, France
\item[$^2$] D\'epartement de math\'ematiques, Universit\'e de Paris XI, 91405 Orsay Cedex, France
\item[] \emph{E-mails:} srichard@math.univ-lyon1.fr, rafael.tiedra@math.u-psud.fr
\end{itemize}}
\end{quote}}
\maketitle


\begin{abstract}
We carry out the spectral analysis of singular matrix valued
perturbations of $3$-dimensional Dirac operators with variable
magnetic field of constant direction. Under suitable assumptions on
the magnetic field and on the perturbations, we obtain a limiting
absorption principle, we prove the absence of singular continuous
spectrum in certain intervals and state properties of the point
spectrum. Constant, periodic as well as diverging magnetic fields
are covered, and Coulomb potentials up to the physical nuclear
charge $Z<137$ are allowed. The importance of an internal-type
operator (a $2$-dimensional Dirac operator) is also revealed in our
study. The proofs rely on commutator methods.
\end{abstract}

\section{Introduction and main results}\label{S1}
\setcounter{equation}{0}

In an earlier paper \cite{RT04} we carried out the spectral analysis
for matrix valued perturbations of three-dimensional Dirac operators
with variable magnetic field of constant direction. Due to some
technical difficulties, two restrictions on the perturbations were
imposed: The perturbations had to be bounded, and the long-range
part had to be of scalar-type. In the present paper both
restrictions are removed. Coulomb potentials up to the physical
nuclear charge $Z<137$ are considered and matrix valued long-range
perturbations are analysed.

When dealing with such a general Coulomb perturbation, one main
difficulty has to be faced: The perturbation is not small with
respect to the unperturbed operator. Therefore most of the usual
technics of perturbation theory are not available and some slightly
more involved tools have to be employed. For instance, without
magnetic field the problem of selfadjointness of Dirac operators
with Coulomb potentials already has a long history. Distinguished
selfadjoint extensions have to be considered, and it took time to
treat the problem up to the coupling constant corresponding to
$Z<137$. We refer for example to the research papers
\cite{Klaus,Nenciu76,Nenciu77} or to the book \cite[Notes
4.3]{Thaller92} for an account on this issue. More recently the
study of Dirac operators with arbitrary Coulomb singularities
was performed in \cite{Georgescu/Mantoiu} and \cite{Xia}.

On the other side the same situation with a magnetic field has been
much less studied. Some results on the spectrum of Dirac operators
with magnetic fields are available for example in
\cite{BS,D,HG,HNW,Y}, but none of these papers deals with very
general magnetic fields and with Coulomb-type singularities. Note
however that some information on selfadjointness for these operators
can be extracted from \cite{Boutet/Purice} and \cite{Chernoff77},
but in these papers the nature of the spectrum is not considered.
The purpose of the present article is to fill in this gap in a
general situation that we shall now describe.

We consider a relativistic spin-$\frac12$ particle evolving in
$\R^3$ in presence of a variable magnetic field of constant
direction. By virtue of the Maxwell equations, we may assume with no
loss of generality that the magnetic field has the form $\vec
B(x_1,x_2,x_3)=\big(0,0,B(x_1,x_2)\big)$. The unperturbed system is
described in the Hilbert space $\ltwo(\R^3;\C^4)$ by the Dirac
operator
$$
H_0:=\alpha_1\Pi_1+\alpha_2\Pi_2+\alpha_3P_3+\beta m,
$$
where $\beta\equiv\alpha_0,\alpha_1,\alpha_2,\alpha_3$ are the usual
Dirac-Pauli matrices, $m$ is the strictly positive mass of the
particle and $\Pi_j:=-i\partial_j-a_j$ are the generators of the
magnetic translations with a vector potential $\vec
a(x_1,x_2,x_3)=\(a_1(x_1,x_2),a_2(x_1,x_2),0\)$ that satisfies
$B=\partial_1a_2-\partial_2a_1$. Since $a_3 =0$, we write
$P_3:=-i\partial_3$ instead of $\Pi_3$.

In the sequel we study the stability of certain parts of the
spectrum of $H_0$ under a matrix valued perturbation $V$. If $V$
satisfies the natural hypotheses introduced below (which allow
Coulomb singularities), and if $H$ is the suitably defined
selfadjoint operator associated with the formal sum $H_0+V$, then we
prove a limiting absorption principle and state properties of the
point spectrum of $H$ in intervals of $\R$ corresponding to gaps in
the symmetrized spectrum of the operator
$H^0:=\sigma_1\Pi_1+\sigma_2\Pi_2+\sigma_3m$ in $\ltwo(\R^2;\C^2)$.
The matrices $\sigma_j$ are the Pauli matrices and the symmetrized
spectrum $\sigma^0_{\rm sym}$ of $H^0$ is the union of the spectra
of $H^0$ and $-H^0$. We stress that our analysis does not require
any restriction on the behaviour of the magnetic field at infinity.
Nevertheless, the pertinence of our work depends on a certain
property of the internal-type operator $H^0$; namely, the size and
the number of gaps in $\sigma^0_{\rm sym}$. For example, in the
special but important case of a nonzero constant magnetic field
$B_0$, $\sigma_{\rm sym}^0$ is equal to $\{\pm\sqrt{2nB_0+m^2}\mid
n=0,1,2,\ldots\}$, which implies that there are plenty of gaps where
our analysis gives results. We refer to \cite{BS,D,HNW} for various
information on the spectrum of $H^0$, especially in the situations
of physical interest, for example when $B$ is constant, periodic or
diverges at infinity. Let us also note that since Coulomb potentials
are allowed in our approach, a more realistic study of Zeeman effect
\cite{HG} is at hand.

In order to state precisely our results, let us introduce some
notations. $\B_{\rm h}(\C^4)$ stands for the set of $4\times4$
hermitian matrices, and $\|\cdot\|$ denotes the norm of the Hilbert
space $\H:=\ltwo(\R^3;\C^4)$ as well as the norm of $\B(\H)$, the
set of bounded linear operators on $\H$. $P_3$ is considered as an
operator in $\H$ or in $\ltwo(\R)$ depending on the context.
$\N:=\{0,1,2,\ldots\}$ is the set of natural numbers. $\vartheta$ is
an arbitrary $C^\infty\big([0,\infty)\big)$-function such that
$\vartheta=0$ near $0$ and $\vartheta=1$ near infinity. $Q_j$ is the
multiplication operator by the coordinate $x_j$ in $\H$, and
$Q:=(Q_1,Q_2,Q_3)$. The notation \ae stands for ``almost everywhere"
and refers to the Lebesgue measure, and the expression
$\<\:\!\cdot\:\!\>$ corresponds to $\sqrt{1+(\:\!\cdot\:\!)^2}$. We
write $\dom(S)$ for the domain of a selfadjoint operator (or a
form) $S$. Finally, the limiting absorption principle for $H$ is
going to be expressed in terms of the Banach space
$\K:=\big(\dom(\<Q_3\>),\H\big)_{1/2,1}$ defined by real
interpolation \cite[Chap. 2]{ABG}. For convenience, we recall that
the weighted space $\H_s:=\dom(\<Q_3\>^s)$ is contained in $\K$ for
each $s>1/2$.

The perturbation $V$ splits into two parts: a regular matrix valued
function and a singular matrix valued function with compact support.
The following definitions concern the former part.

\begin{Definition}\label{defdespots}
Let $V$ be a multiplication operator associated with an element of
$\linf\big(\R^3;\B_{\rm h}(\C^4)\big)$.
\begin{itemize}
\item[(a)] $V$ is \:\!\emph{small at infinity}\:\! if~
$\lim_{r\to\infty}\left\|\vartheta(|Q|/r)V\right\|=0$,
\item[(b)] $V$ is \:\!\emph{short-range}\:\! if~
$\int_1^\infty\d r\left\|\vartheta(|Q_3|/r)V\right\|<\infty$,
\item[(c)] Assume that $V$ is continuously differentiable with respect to $x_3$, and that the map
$x\mapsto\<x_3\>(\partial_3V)(x)$ belongs to $\linf\big(\R^3;\B_{\rm
h}(\C^4)\big)$, then $V$ is \:\!\emph{long-range}\:\! if
$$
\int_1^{\infty}\frac{\d
r}r\left\|\vartheta(|Q_3|/r)\<Q_3\>(\partial_3 V)\right\|<\infty.
$$
\end{itemize}
\end{Definition}

Note that Definitions \ref{defdespots}.(b) and \ref{defdespots}.(c)
differ from the standard ones: the decay rate is imposed only in the
$x_3$ direction. In the sequel we consider a magnetic field
$B\in\linfloc(\R^2;\R)$ and always choose a vector potential $\vec
a=(a_1,a_2,0)$ in $\linfloc(\R^2;\R^3)$, \eg the one obtained by
means of the transversal gauge \cite[Sec.~8.4.2]{Thaller92}. We are
now in a position to state our main result. Let us already mention
that Proposition \ref{autoadjonction} contains more information on
the distinguished selfadjoint operator $H$.

\begin{Theorem}\label{MainThm}
Assume that $B$ belongs to $\linfloc(\R^2;\R)$ and that $V(x)$
belongs to $\B_{\rm h}(\C^4)$ for a.e. $x\in\R^3$. Suppose that
there exist $\chi\in C_0^\infty(\R^3;\R)$, a finite set
$\Gamma\subset\R^3$, and a positive number $\nu<1$ such that:
\begin{enumerate}
\item[(i)] $V_{\rm reg}:=(1-\chi)V$ belongs to $\linf\big(\R^3;\B_{\rm h}(\C^4)\big)$, is small
at infinity and can be written as the sum of a short-range and a
long-range potential,
\item[(ii)] $V_{\rm sing}:=\chi V$ can be written as the sum of two matrix valued Borel functions
$V_{\sf loc}\in\lthreeloc\big(\R^3;\B_{\rm h}(\C^4)\big)$ and
$V_{\sf c}$ with
$$
\left\|V_{\sf c}(x)\right\|_{\B_{\rm
h}(\C^4)}\le\sum_{a\in\Gamma}\frac\nu{|x-a|} \quad\forall x\in\R^3.
$$
\end{enumerate}
Then there exists a unique selfadjoint operator $H$ in $\H$,
formally equal to $H_0+V$, with domain $\dom(H)\subset\H_{\rm
loc}^{1/2}(\R^3;\C^4)$, such that
\begin{enumerate}
\item[(a)] $\sigma_{\rm ess}(H)=\sigma_{\rm ess}(H_0)$,
\item[(b)] The point spectrum of the operator $H$ in $\R\setminus\sigma^0_{\rm sym}$ is
composed of eigenvalues of finite multiplicity and with no
accumulation point in $\R\setminus\sigma^0_{\rm sym}$,
\item[(c)] The operator $H$ has no singular continuous spectrum in
$\R\setminus\sigma^0_{\rm sym}$,
\item[(d)] The limits $\lim_{\varepsilon\searrow0}\<\psi,(H-\lambda\mp i\varepsilon)^{-1}\psi\>$
exist for each $\psi\in\K$, uniformly in $\lambda$ on each compact
subset of $\R\setminus\{\sigma_{\rm sym}^0\cup\sigma_{\rm
pp}(H)\}$.\end{enumerate}
\end{Theorem}

As usual, the limiting absorption principle obtained in (d) leads to
$H$-smooth operators, which imply for suitable short-range
perturbations the existence of local wave operators. Since these
constructions are rather standard, we shall not develop them here.

Let us finally give a description of the organisation of this paper
and make a comment on its relation with the earlier work
\cite{RT04}. In Section \ref{surlopH_0} we study the operator $H_0$
and construct a suitable operator conjugated to $H_0$. The Mourre
estimate is given at the end of Section \ref{sec22}. Regular
perturbations are introduced in Section \ref{perturbornee} and their
properties with respect to the conjugate operator are then obtained.
A version of Theorem \ref{MainThm} for regular perturbations is
proved in Theorem \ref{thmborne}. In Section \ref{SectionSingular}
the singular part of the potential is added and a description of the
selfadjoint operator $H_0+V$ is given in Proposition
\ref{autoadjonction}. Last part of Section \ref{SectionSingular} is
devoted to the proof of our main result in its full generality.

The major improvements contained in this paper are mainly due to (i)
the use of a simple scalar conjugate operator (see Section
\ref{sec22}), and (ii) the application of the new approach of
\cite{Georgescu/Mantoiu} developed for dealing with singular
perturbations (see Section \ref{SectionSingular}). These two new
technical tools allow us to treat Coulomb singularities and
long-range matrix valued potentials. In the same time, we extend the
class of magnetic fields that can be considered from continuous ones
to locally bounded ones. Due to these various improvements, not a
single result from \cite{RT04} can be quoted without changing its
statement or its proof. Therefore the present paper is
self-contained and does not depend on any previous results from
\cite{RT04}.

\section{The unperturbed operator}\label{surlopH_0}
\setcounter{equation}{0}

\subsection{Preliminaries}\label{prelim}

Let us start by recalling some known results. Since $\vec a$ belongs
to $\linfloc(\R^2;\R^3)$, it follows from \cite[Lem.
4.3]{Georgescu/Mantoiu} and \cite[Thm. 1.3]{Boutet/Purice} that
$H_0$ is essentially selfadjoint on $\D:=C^\infty_0(\R^3;\C^4)$,
with $\dom(H_0)\subset \H^{1/2}_{\rm loc} \equiv\H^{1/2}_{\rm
loc}(\R^3;\C^4)$ (the local Sobolev space of order $1/2$ of
functions on $\R^3$ with values in $\C^4$). Moreover the spectrum of
$H_0$  is symmetric with respect to $0$ and does not contain the
interval $(-m,m)$ \cite[Sec. 5.5.2\;\&\;Cor. 5.14]{Thaller92}.

We now introduce a suitable representation of the Hilbert space
$\H$. We consider the partial Fourier transformation
\begin{equation}\label{F unitaire}
\F:\D\to\int_\R^\oplus\d\xi\,\H_{\scriptscriptstyle 12},
\qquad(\F\psi)(\xi):=\frac1{\sqrt{2\pi}}\int_\R\d x_3\,\e^{-i\xi
x_3}\psi(\:\!\cdot\:\!,x_3),
\end{equation}
where $\H_{\scriptscriptstyle 12}:=\ltwo(\R^2;\C^4)$. This map
extends uniquely to a unitary operator from $\H$ onto
$\int_\R^\oplus\d\xi\,\H_{\scriptscriptstyle 12}$, which we denote
by the same symbol $\F$. One obtains then the following direct
integral decomposition of $H_0$:
$$
\F H_0\F^{-1}=\int_\R^\oplus\d\xi\,H_0(\xi),
$$
where $H_0(\xi)$ is the selfadjoint operator in
$\H_{\scriptscriptstyle 12}$ acting as
$\alpha_1\Pi_1+\alpha_2\Pi_2+\alpha_3\xi+\beta m$ on
$C_0^\infty(\R^2;\C^4)$. In the following remark we draw the
connection between the operators $H_0$, $H_0(0)$ and the
internal-type operator $H^0$.

\begin{Remark}\label{jubajuba}
The operator $H_0(0)$ acting on $C_0^{\infty}(\R^2;\C^4)$ is
unitarily equivalent to the direct sum $ \footnotesize
\begin{pmatrix}
m & \Pi_-\\
\Pi_+ & -m
\end{pmatrix}
\oplus
\begin{pmatrix}
m & \Pi_+
\\ \Pi_- & -m
\end{pmatrix}
$ acting on $C_0^\infty(\R^2;\C^2)\oplus C_0^\infty(\R^2;\C^2)$,
where $\Pi_\pm:=\Pi_1\pm i\Pi_2$. These two matrix operators act in
$\ltwo(\R^2;\C^2)$ and are essentially selfadjoint on
$C_0^\infty(\R^2;\C^2)$ \cite[Thm.~2.1]{Chernoff77}. The first one
is equal to $H^0$, while the second one is unitarily equivalent to
$-H^0$ (this can be obtained with the abstract Foldy-Wouthuysen
transformation \cite[Thm.~5.13]{Thaller92}). Therefore $H_0(0)$ is
essentially selfadjoint on $C_0^{\infty}(\R^2;\C^4)$, and
$H_0(\xi)=H_0(0)+\alpha_3\xi$ for each $\xi\in\R$. Since
$\alpha_3H_0(0)+H_0(0)\alpha_3=0$ it follows that
$H_0(\xi)^2=H_0(0)^2+\xi^2$, and
\begin{equation}\label{brouetteandalouse}
\sigma[H_0(\xi)^2]=\sigma[H_0(0)^2+\xi^2]=(\sigma_0^{\rm
sym})^2+\xi^2.
\end{equation}
Thus one has the identity
$$
H_0^2=H_0(0)^2\otimes1+1\otimes P_3^2
$$
with respect to the tensorial decomposition
$\ltwo(\R^2;\C^4)\otimes\ltwo(\R)$ of $\H$. In particular the
spectrum of $H_0^2$ is purely absolutely continuous and equal to the
interval $[\mu_0^2,\infty)$, where $\mu_0:=\inf|\sigma^0_{\rm
sym}|\ge m$. Since the spectrum of $H_0$ is symmetric with respect
to $0$, it follows that
$$
\sigma(H_0)=(-\infty,-\mu_0]\cup[\mu_0,+\infty).
$$
\end{Remark}

We now state three technical lemmas that are constantly used in the
sequel. Proofs can be found in the appendix.

\begin{Lemma}\label{technique1}
\begin{enumerate}
\item[(a)] For each $n\in \N$, $H_0^{-n}\D$ and $|H_0|^{-n}\D$  are included in $\dom(Q_3)$,
\item[(b)] $P_3 |H_0|^{-1}$ is a bounded selfadjoint operator equal to $|H_0|^{-1}P_3$ on
$\dom(P_3)$. In particular, $|H_0|^{-1}\H$ is included in
$\dom(P_3)$.
\end{enumerate}
\end{Lemma}

\begin{Lemma}\label{technique2}
Let $g$ be in $C^1(\R)$ with $g'$ bounded, and let $n\in\N$. Then
$\dom(Q_3)$ is included in $\dom[g(Q_3)]$, and the following
equality holds on $H_0^{-n}\D$:
\begin{equation*}
H_0^{-1}g(Q_3)-g(Q_3)H_0^{-1}=iH_0^{-1}\alpha_3g'(Q_3)H_0^{-1}.
\end{equation*}
\end{Lemma}

The last statement implies that the commutator of $H_0^{-1}$ and
$g(Q_3)$, defined on the core $\D$ of $g(Q_3)$, extends uniquely
to a bounded operator. In the framework of \cite[Def.~6.2.2]{ABG},
this means that the operator $H_0$ is of class
$C^1\big(g(Q_3)\big)$.

Given two appropriate functions $f$ and $g$, we recall some
properties of the commutator $[f(P_3),g(Q_3)]$ acting in the
weighted space $\H_s$, $s\in\R$. We use the notation $\widehat f$
for the Fourier transform of $f$, and $S^m(\R)$ for the vector space
of symbols of degree $m$ on $\R$.

\begin{Lemma}\label{technique3}
Let $s\geq 0$ and $g\in S^1(\R)$. Suppose that $f\in BC^\infty(\R)$
is such that $x\mapsto\<x\>^s\widehat{f'}(x)$ belongs to
$\lone(\R)$. Then $f(P_3)$ leaves $\dom(Q_3)$ invariant, and the
operator $f(P_3)g(Q_3)-g(Q_3)f(P_3)$ defined on $\dom(Q_3)$ extends
uniquely to an operator in $\B(\H)$, which is denoted by
$[f(P_3),g(Q_3)]$. Furthermore, this operator restricts to an
element of $\B(\H_s)$.
\end{Lemma}

\subsection{Strict Mourre estimate for the free Hamiltonian}\label{sec22}

We now gather some results on the regularity of $H_0$ with respect
to a conjugate operator. This operator is constructed with a
function $F$ satisfying the following hypotheses.

\begin{Assumption}\label{ElChichi}
$F$ is a non-decreasing element of $C^\infty(\R;\R)$ with $F(x)=0$
for $x\leq 0$ and $F(x)=1$ for $x \geq 1$.
\end{Assumption}

A useful property of such a function is that $\widehat{F^{(k)}}$
belongs to the Schwartz space on $\R$, for any integer $k>0$. In the
sequel we always assume that $F$ is a function satisfying Assumption
\ref{ElChichi}. In particular, it follows that the formal expression
\begin{equation}\label{defdeA}
A:=\mbox{$\frac12$}[Q_3F(P_3)+F(P_3)Q_3]
\end{equation}
leads to a well-defined symmetric operator on $\D$.

\begin{Lemma}\label{sur A}
The operator $A$ is essentially selfadjoint on $\D$, and its closure
is essentially selfadjoint on any core for $\<Q_3\>$.
\end{Lemma}

\begin{proof}
The claim is a consequence of Nelson's criterion of essential
selfadjointness \cite[Thm.~X.37]{RS} applied to the triple
$\{\<Q_3\>,A,\D\}$. So we simply verify the two hypotheses of that
theorem. By using Lemma \ref{technique3}, one first obtains that for
all $\psi\in\D$:
$$
\|A\psi\|=\left\|\(F(P_3)Q_3-\mbox{$\frac12$}
\[F(P_3), Q_3\]\)\psi\right\|\le\textsc c\left\|\< Q_3\>\psi\right\|
$$
for some constant $\textsc c>0$ independent of $\psi$. Furthermore,
one has for all $\psi\in\D$:
$$
\<A\psi,\<Q_3\>\psi\>-\<\<Q_3\>\psi,A\psi\>
=\mbox{$\frac12$}\left\{\<Q_3\psi,[F(P_3),\<Q_3\>]\psi\>
-\<[F(P_3),\<Q_3\>]\psi,Q_3\psi\>\right\}.
$$
Since $[F(P_3),\<Q_3\>]\in\B(\H_{1/2})$ by Lemma \ref{technique3}
and since $Q_3\in\B(\H_{1/2},\H_{-1/2})$, one easily gets the
estimate
$$
|\<A\psi,\<Q_3\>\psi\>-\<\<Q_3\>\psi,A\psi\>|\le\textsc
d\!\;\big\|\<Q_3\>^{1/2}\psi\big\|^2
$$
for all $\psi\in\D$ and a constant $\textsc d>0$ independent of
$\psi$.
\end{proof}

>From now on we set $\G:=\dom(H_0)$, and we write $\G^*$ for the
adjoint space of $\G$. One has the continuous dense embeddings
$\G\hookrightarrow\H\hookrightarrow\G^*$, where $\H$ is identified
with its adjoint through the Riesz isomorphism. In the sequel we
constantly use the fact that the bounded operators $H_0^{-1}$ and
$F(P_3)$ commute.

\begin{Proposition}\label{regulariteH0}
\begin{enumerate}
\item[(a)] The quadratic form
$\dom(A)\owns\psi\mapsto\<H_0^{-1}\psi,iA\psi\>-\<A\psi,iH_0^{-1}\psi\>$
extends uniquely to the bounded form defined by the operator
$-H_0^{-1}\alpha_3F(P_3)H_0^{-1}\in\B(\H)$.
\item[(b)] The group $\{e^{itA}\}_{t\in\R}$ leaves $\G$ invariant.
\item[(c)] The quadratic form
\begin{equation}\label{form1}
\dom(A)\ni\psi\mapsto\<H_0^{-1}\alpha_3F(P_3)H_0^{-1}\psi,iA\psi\>
-\<A\psi,iH_0^{-1}\alpha_3F(P_3)H_0^{-1}\psi\>,
\end{equation}
extends uniquely to a bounded form on $\H$.
\end{enumerate}
\end{Proposition}

In the framework of \cite[Def.~6.2.2]{ABG}, statements (a) and (c)
mean that $H_0$ is of class $C^1(A)$ and $C^2(A)$ respectively.

\begin{proof}
(a) For any $\psi\in\D$, one gets
\begin{align}
&2\[\<H_0^{-1}\psi,iA\psi\>-\<A\psi,iH_0^{-1}\psi\>\]\nonumber\\
&=\<i[H_0^{-1},Q_3]\psi,F(P_3)\psi\>+\<F(P_3)\psi,i[H_0^{-1},Q_3]\psi\>\nonumber\\
&=-2\<\psi,H_0^{-1}\alpha_3F(P_3)H_0^{-1}\psi\>,\label{first
commutator}
\end{align}
by using Lemma \ref{technique2}. Since $\D$ is a core for $A$, then
the statement follows by density. We shall write $i[H_0^{-1},A]$ for
the bounded extension of the quadratic form
$\dom(A)\owns\psi\mapsto\<H_0^{-1}\psi,iA\psi\>-\<A\psi,iH_0^{-1}\psi\>$.

(b) Let $i[H_0,A]$ be the operator in $\B(\G,\G^*)$ associated with
the unique extension to $\G$ of the quadratic form
$\psi\mapsto\<H_0\psi,iA\psi\>-\<A\psi,iH_0\psi\>$ defined for all
$\psi\in\G\cap\dom(A)$. Then $\G$ is invariant under
$\{e^{itA}\}_{t\in\R}$ if $H_0$ is of class $C^1(A)$ and if
$i[H_0,A]\G\subset\H$ \cite[Lemma 2]{Georgescu/Gerard}. From
Equation \eqref{first commutator} and \cite[Eq.~6.2.24]{ABG}, one
obtains the following equalities valid in form sense on $\H$:
$$
-H_0^{-1}\alpha_3F(P_3)H_0^{-1}=i[H_0^{-1},A]=-H_0^{-1}i[H_0,A]H_0^{-1}.
$$
Thus $i[H_0,A]$ and $\alpha_3F(P_3)$ are equal as operators in
$\B(\G,\G^*)$. But since the latter applies $\G$ into $\H$,
$i[H_0,A]\G$ is included in $\H$.

(c) The boundedness on $\D$ of the quadratic form \eqref{form1}
follows by inserting \eqref{defdeA} into the r.h.s. term of
\eqref{form1}, by applying repeatedly Lemma \ref{technique2} with
$g(Q_3)=Q_3$, and by taking Lemma \ref{technique3} into account.
Then one concludes by using the density of $\D$ in $\dom(A)$.
\end{proof}

It will also be useful to show that $|H_0|$ is of class $C^1(A)$.

\begin{Lemma}\label{jaimalaupoignet}
The quadratic form
$\dom(A)\owns\psi\mapsto\<|H_0|^{-1}\psi,iA\psi\>-\<A\psi,i|H_0|^{-1}\psi\>$
extends uniquely to the bounded form defined by
$-|H_0|^{-1}F(P_3)P_3|H_0|^{-2}\in\B(\H)$.
\end{Lemma}

\begin{proof}
A direct calculation using the transformation \eqref{F unitaire} and
Lemma \ref{technique1} gives for any $\psi\in\D$
$$
i[|H_0|^{-1},Q_3]\psi=-P_3|H_0|^{-3}\psi.
$$
Thus one has the equalities
\begin{align*}
&2\(\<|H_0|^{-1}\psi,iA\psi\>-\<A\psi,i|H_0|^{-1}\psi\>\)\\
&=\<i[|H_0|^{-1},Q_3]\psi,F(P_3)\psi\>+\<F(P_3)\psi,i[|H_0|^{-1},Q_3]\psi\>\\
&=-2\<\psi,|H_0|^{-1}F(P_3)P_3|H_0|^{-2}\psi\>.
\end{align*}
Since $\D$ is a core for $A$, then the statement follows by density.
\end{proof}

Due to Lemma \ref{jaimalaupoignet} and \cite[Eq.~6.2.24]{ABG} the
operator $i[|H_0|,A]$ associated with the unique extension to $\G$
of the quadratic form
$\G\cap\dom(A)\ni\psi\mapsto\<|H_0|\psi,iA\psi\>-\<A\psi,i|H_0|\psi\>$
is equal to $F(P_3)P_3|H_0|^{-1}\in\B(\H)$. From now on we simply
write $R$ for this operator and $T$ for the operator
$\alpha_3F(P_3)\equiv i[H_0,A]\in\B(\H)$.

In the following definition, we introduce two functions giving the
optimal value to a Mourre-type inequality. Remark that a slight
modification has been done with regard to the usual definition
\cite[Eq.~7.2.4]{ABG}.

\begin{Definition}
Let $H$ be a selfadjoint operator in a Hilbert space $\H$ and assume
that $S$ is a symmetric operator in $\B\big(\dom(H),\dom(H)^*\big)$.
Let
$E^H(\lambda;\varepsilon):=E^H\big((\lambda-\varepsilon,\lambda+\varepsilon)\big)$
be the spectral projection of $H$ for the interval
$(\lambda-\varepsilon,\lambda+\varepsilon)$. Then, for all
$\lambda\in\R$ and $\varepsilon>0$, we set
\begin{align*}
\varrho^S_H(\lambda;\varepsilon)&:=\sup\left\{a\in\R\mid
E^H(\lambda;\varepsilon)SE^H(\lambda;\varepsilon)\ge a\:\!E^H(\lambda;\varepsilon)\right\},\\
\varrho^S_H(\lambda)&:=\sup_{\varepsilon>0}\varrho^S_H(\lambda;\varepsilon).
\end{align*}
\end{Definition}

Let us make three observations: the inequality
$\varrho_H^S(\lambda;\varepsilon')\le\varrho_H^S(\lambda;\varepsilon)$
holds whenever $\varepsilon'\ge\varepsilon$,
$\varrho^S_H(\lambda)=+\infty$ if $\lambda$ does not belong to the
spectrum of $H$, and $\varrho^S_H(\lambda)\ge0$ for all
$\lambda\in\R$ if $S\ge0$. We also mention that in the case of two
selfadjoint operators $H$ and $A$ in $\H$, with $H$ of class
$C^1(A)$ and $S=i[H,A]$, the function $\varrho_H^S(\:\!\cdot\:\!)$
is equal to the function $\varrho_H^A(\:\!\cdot\:\!)$ defined in
\cite[Eq.~7.2.4]{ABG}.

\begin{Lemma}\label{rhoegalite}
For $\lambda>0$ and $\varepsilon\in(0,\lambda)$, one has
$\varrho_{H_0}^T(\lambda;\varepsilon)=\varrho_{H_0}^R(\lambda;\varepsilon)$.
Similarly, for $\lambda<0$ and $\varepsilon\in (0,|\lambda|)$, one
has
$\varrho_{H_0}^{-T}(\lambda;\varepsilon)=\varrho_{H_0}^R(\lambda;\varepsilon)$.
\end{Lemma}

\begin{proof}
We give the proof of the first equality, the second one can be
obtained in the same way.

Let $\varphi\in C^\infty_0(\R;\R)$ with
$\supp(\varphi)\subset(0,\infty)$, and let $\psi\in\dom(A)$. Since
$\varphi(H_0)\in C^1(A)$ \cite[Thm.~6.2.5]{ABG}, then
$\varphi(H_0)\psi\in\G \cap\dom(A)$. Thus by using the spectral
theorem we get
\begin{align*}
&\<\psi,\varphi(H_0)T\varphi(H_0)\psi\>\\
&=\<H_0\varphi(H_0)\psi,iA\varphi(H_0)\psi\>-\<A\varphi(H_0)\psi,iH_0\varphi(H_0)\psi\>\\
&=\<|H_0|\varphi(H_0)\psi,iA\varphi(H_0)\psi\>-\<A\varphi(H_0)\psi,i|H_0|\varphi(H_0)\psi\>\\
&=\<\psi,\varphi(H_0)R\varphi(H_0)\psi\>.
\end{align*}
Since $\dom(A)$ is dense in $\H$ the identity
$$
\<\psi,\varphi(H_0)T\varphi(H_0)\psi\>
=\<\psi,\varphi(H_0)R\varphi(H_0)\psi\>
$$
even holds for each $\psi\in\H$. Now, for $\lambda>0$ and
$\varepsilon\in(0,\lambda)$ fixed one may choose $\eta\in
C^\infty_0(\R;\R)$ with $\supp(\eta)\subset(0,\infty)$ satisfying
$\eta(x)=1$ for all $x\in[\lambda-\varepsilon,\lambda+\varepsilon]$.
Then
\begin{align*}
E^{H_0}(\lambda;\varepsilon)TE^{H_0}(\lambda;\varepsilon)
&= E^{H_0}(\lambda;\varepsilon)\eta(H_0)T\eta(H_0)E^{H_0}(\lambda;\varepsilon)\\
&=E^{H_0}(\lambda;\varepsilon)\eta(H_0)R\eta(H_0)E^{H_0}(\lambda;\varepsilon) \\
&=E^{H_0}(\lambda;\varepsilon)RE^{H_0}(\lambda;\varepsilon),
\end{align*}
and the proof is complete.
\end{proof}

The operator $\F R\F^{-1}$ is decomposable, more precisely:
$$
\F R\F^{-1}=\int_\R^\oplus\d\xi\,R(\xi)\qquad{\rm with}\qquad
R(\xi)=F(\xi)\xi|H_0(\xi)|^{-1}\in\B(\H_{\scriptscriptstyle 12}).
$$
Taking advantage of this and of the direct integral decomposition of
$H_0$, one obtains for each $\lambda\in\R$ and $\varepsilon >0$ the
formula
\begin{equation}\label{unmercredimatin}
\varrho^R_{H_0}(\lambda;\varepsilon)
=\essinf_{\xi\in\R}\varrho^{R(\xi)}_{H_0(\xi)}(\lambda;\varepsilon).
\end{equation}

Now we can deduce a lower bound for
$\varrho_{H_0}^T(\:\!\cdot\:\!)$.

\begin{Proposition}\label{Mourre for H_0}
For $\lambda\ge0$ one has
\begin{equation}\label{unlundimatin}
\varrho_{H_0}^T(\lambda)\ge\inf
\bigg\{\frac{F\big(\sqrt{\lambda^2-\mu^2}\big)\sqrt{\lambda^2-\mu^2}}\lambda
\mid\mu\in\sigma_{\rm sym}^0\cap\[0,\lambda\]\bigg\}
\end{equation}
with the convention that the infimum over an empty set is $+\infty$.
\end{Proposition}

\begin{proof}
(i) Recall from Remark \ref{jubajuba} that $\mu_0=\inf|\sigma^0_{\rm
sym}|=\inf\{\sigma(H_0)\cap [0,+\infty)\}$. Thus, for
$\lambda\in[0,\mu_0)$ the l.h.s. of \eqref{unlundimatin} is equal to
$+\infty$, since $\lambda$ does not belong to the spectrum of $H_0$.
Then, \eqref{unlundimatin} is obviously satisfied on $[0,\mu_0)$.

(ii) If $\lambda\in\sigma^0_{\rm sym}$, then the r.h.s. term of
\eqref{unlundimatin} is equal to $0$. However, due to Lemma
\ref{rhoegalite} and the positivity of $R$, we have
$\varrho^T_{H_0}(\lambda)\ge0$. Hence the relation
\eqref{unlundimatin} is again satisfied.

(iii) Let $0<\varepsilon<\mu_0<\lambda$. Direct computations using
the explicit form of $R(\xi)$ and the spectral theorem for the
operator $H_0(\xi)$ show that for $\xi$ fixed, one has
\begin{equation}\label{verre}
\varrho^{R(\xi)}_{H_0(\xi)}(\lambda;\varepsilon)
=\inf\bigg\{\frac{F(\xi)\xi}{|\rho|}\mid\rho\in(\lambda
-\varepsilon,\lambda +\varepsilon)
\cap\sigma[H_0(\xi)]\bigg\}\ge\frac{F(\xi)\xi}{\lambda +
\varepsilon}\ .
\end{equation}
On the other hand one has
$\varrho^{R(\xi)}_{H_0(\xi)}(\lambda;\varepsilon)=+\infty$ if
$(\lambda -\varepsilon,\lambda
+\varepsilon)\cap\sigma[H_0(\xi)]=\varnothing$, and a fortiori
$$
\varrho^{R(\xi)}_{H_0(\xi)}(\lambda;\varepsilon)=+\infty\qquad\textrm{if}\qquad
\big((\lambda -\varepsilon)^2,(\lambda
+\varepsilon)^2\big)\cap\sigma[H_0(\xi)^2]=\varnothing.
$$
Thus, by taking into account Equations \eqref{unmercredimatin},
\eqref{verre}, the previous observation and relation
\eqref{brouetteandalouse}, one obtains that
\begin{equation}\label{estimation1}
\varrho^R_{H_0}(\lambda;\varepsilon)\ge\essinf
\bigg\{\frac{F(\xi)\xi}{\lambda + \varepsilon}\mid\xi^2\in
\big((\lambda -\varepsilon)^2, (\lambda +\varepsilon)^2\big)
-(\sigma^0_{\rm sym})^2\bigg\}.
\end{equation}
Suppose now that $\lambda\not\in\sigma^0_{\rm sym}$, define
$\mu:=\sup\{\sigma^0_{\rm sym}\cap[0,\lambda]\}$ and choose
$\varepsilon>0$ such that $\mu<\lambda-\varepsilon$. Then the
inequality \eqref{estimation1} implies that
$$
\varrho^R_{H_0}(\lambda;\varepsilon) \ge\frac{F\big(\sqrt{(\lambda -
\varepsilon)^2-\mu^2}\big) \sqrt{(\lambda -
\varepsilon)^2-\mu^2}}{\lambda + \varepsilon}.
$$
Since
$\varrho_{H_0}^T(\lambda;\varepsilon)=\varrho_{H_0}^R(\lambda;\varepsilon)$,
the relation \eqref{unlundimatin} follows from the above formula
when $\varepsilon\searrow0$.
\end{proof}

\begin{Remark}\label{FishAndChips}
Using the conjugate operator $-A$ instead of $A$, and thus dealing
with $-T$ instead of $T$, one can show as in Proposition \ref{Mourre
for H_0} that $-A$ is strictly conjugate to $H_0$ on $(-\infty,
0] \setminus \sigma^0_{\rm sym}$; more precisely, one has for each
$\lambda\le0$
\begin{equation}\label{jeudipipi}
\varrho_{H_0}^{-T}(\lambda)\ge\inf
\bigg\{\frac{F\big(\sqrt{\lambda^2-\mu^2}\big)\sqrt{\lambda^2-\mu^2}}{|\lambda|}
\mid\mu\in\sigma_{\rm sym}^0\cap\[0,|\lambda|\]\bigg\},
\end{equation}
with the convention that the infimum over an empty set is $+\infty$.
In the rest of the paper, for the sake of brevity, we shall mostly
concentrate on the positive part of the spectrum of $H_0$, and give
few hints on the trivial adaptations for the negative part of the
spectrum.
\end{Remark}

\section{The bounded perturbation}\label{perturbornee}
\setcounter{equation}{0}

In this section we consider the operator $H:=H_0+W$ with a potential
$W$ belonging to $\linf\big(\R^3;\B_{\rm h}(\C^4)\big)$. The
operator $H$ is selfadjoint and its domain is equal to the domain
$\G\equiv\dom(H_0)$ of $H_0$. We first give a result on the
difference of the resolvents $(H-z)^{-1}-(H_0-z)^{-1}$ and, as a
corollary, we obtain the localization of the essential spectrum of
$H$. For that purpose we recall that a selfadjoint operator $S$ in
$\H$ is said to be locally compact if $\eta(Q)(S+i)^{-1}$ is a
compact operator for each $\eta\in C_0(\R^3)$.

\begin{Lemma}\label{compactness}
Assume that $W$ is small at infinity. Then for all
$z\in\C\setminus\{\sigma(H)\cup\sigma(H_0)\}$ the difference
$(H-z)^{-1}-(H_0-z)^{-1}$ is a compact operator. In particular
$\sigma_{\rm ess}(H)=\sigma_{\rm ess}(H_0)$.
\end{Lemma}

\begin{proof}
Since $W$ is bounded and small at infinity, it is enough to check
that $H_0$ is locally compact \cite[Sec.~4.3.4]{Thaller92}. However,
as already mentioned in Section \ref{prelim}, one has
$\G\subset\H^{1/2}_{\rm loc}$. Hence the statement follows by usual
arguments.
\end{proof}

In order to obtain a limiting absorption principle for $H$, we shall
invoke some abstract results. For that purpose, we first prove an
optimal regularity condition of $H$ with respect to $A$. We refer to
\cite[Chap.~5]{ABG} for the definitions of the classes $C^{1,1}(A)$
and $C^{1,1}(A;\G,\G^*)$, and for more explanations on regularity
conditions. The optimality of the regularity condition in the
framework of commutator methods is shown in \cite[App.~7.B]{ABG}.

\begin{Proposition}\label{V regularity}
Let $W$ be the sum of a short-range and a long-range potential. Then
$H=H_0+W$ is of class $C^{1,1}(A)$.
\end{Proposition}

\begin{proof}
Since $\{e^{itA}\}_{t\in\R}$ leaves $\dom(H)\equiv\G$ invariant, it
is equivalent to prove that $H$ belongs to $C^{1,1}(A;\G,\G^*)$
\cite[Thm.~6.3.4.(b)]{ABG}. But in Proposition
\ref{regulariteH0}.(c), it has already been shown that $H_0$ is of
class $C^2(A)$, so that $H_0$ belongs to $C^{1,1}(A;\G,\G^*)$. Thus
it is enough to prove that $W$ belongs to $C^{1,1}(A;\G,\G^*)$,
which is readily satisfied if $W \in C^{1,1}(A)$. In the short-range
case, we shall use \cite[Thm.~7.5.8]{ABG} for the couple $\H$ and
$\<Q_3\>$. The non-trivial conditions needed for that theorem are
obtained in point (i) below. In the long-range case, the claim
follows by \cite[Thm.~7.5.7]{ABG}, which can be applied because of
point (ii) below.

(i) The first condition is trivially satisfied since $\{e^{it\<
Q_3\>}\}_{t\in\R}$ is a unitary $C_0$-group in $\H$. For the second
condition, one has to check that $\<Q_3\>^{-2}A^2$ defined on
$\dom(A^2)$ extends to an operator in $\B(\H)$. After some
commutator calculations performed on $\D$, one obtains that
$\<Q_3\>^{-1}A$ and $\<Q_3\>^{-2}A$ are respectively equal on $\D$
to some operators $S_1$ and $S_2\<Q_3\>^{-1}$ in $\B(\H)$, where
$S_1$ and $S_2$ are linear combinations of products of operators
$f(P_3)$, $g(G_3)$ and $[h(P_3),\<Q_3\>]$ with $f,g,h\in
BC^\infty(\R;\R)$ and $\widehat{h'}\in\lone(\R)$. Since $\D$ is a
core for $A$, these equalities even hold on $\dom(A)$. Hence one has
on $\dom(A^2)$:
$$
\<Q_3\>^{-2}A^2=\big(\<Q_3\>^{-2}A\big)A=S_2\<Q_3\>^{-1}A=S_2S_1.
$$
In consequence $\<Q_3\>^{-2}A^2$ is equal on $\dom(A^2)$ to an
operator in $\B(\H)$. The statement follows then by density.

(ii) It has been proved in Lemma \ref{sur A} that the inclusion
$\dom(\<Q_3\>)\subset\dom(A)$ holds. Furthermore the inequality
$r\left\|(\< Q_3\>+ir)^{-1}\right\|\le\textrm{Const.}$ for all $r>0$
is trivially satisfied. Thus one is left in proving that the
commutator $i[W,A]$, defined as a quadratic form on $\dom(A)$, with
$W$ a long-range potential, is bounded and satisfies the estimate
$$
\int_1^\infty\frac{\d r}r\left\|\vartheta(\<Q_3\>/r)[W,A]\right\|<\infty
$$
for an arbitrary function $\vartheta \in
C^\infty\big([0,\infty)\big)$ with $\vartheta=0$ near $0$ and
$\vartheta=1$ near infinity. However, such an estimate can be
obtained by mimicking the proof given in \cite[p.~345]{ABG} and by
taking into account the particular properties of $F$.
\end{proof}

\begin{Theorem}\label{thmborne}
Assume that $B$ belongs to $\linfloc(\R^2;\R)$, and that \:\!$W$
belongs to \:\!$\linf\big(\R^3;\mathscr B_{\rm h}(\C^4)\big)$, is
small at infinity and can be written as the sum of a short-range and
a long-range potential. Then statments (a) to (d) of Theorem
\ref{MainThm} hold for $H=H_0+W$.
\end{Theorem}

\begin{proof}
Statement (a) has already been proved in Lemma \ref{compactness}.
Proposition \ref{V regularity} implies that both $H_0$ and $H$ are
of class $C^{1,1}(A)$. Furthermore, the difference
$(H+i)^{-1}-(H_0+i)^{-1}$ is compact by Lemma \ref{compactness}, and
$\varrho^T_{H_0}>0$ on $[0,\infty)\setminus\sigma_{\rm sym}^0$ due
to Proposition \ref{Mourre for H_0}. Hence $A$ is strictly conjugate
to $H$ on $[0,\infty)\setminus\{\sigma_{\rm sym}^0\cup\sigma_{\rm pp}(H)\}$
due to \cite[Thm.~7.2.9\,\&\,Thm.~7.2.13]{ABG}. Similar arguments taking
Remark \ref{FishAndChips} into account show that $-A$ is strictly conjugate
to $H$ on $(-\infty,0]\setminus\{\sigma_{\rm sym}^0\cup\sigma_{\rm
pp}(H)\}$. The assertions (b) and (c) then follow by the abstract
conjugate operator method \cite[Cor.~7.2.11\,\&\,Thm. 7.4.2]{ABG}.

The limiting absorption principle directly obtained via
\cite[Thm.~7.4.1]{ABG} is expressed in terms of the interpolation
space $\big(\dom(A),\H\big)_{1/2,1}$, and of its adjoint. Since both
are not standard spaces, one may use \cite[Cor.~2.6.3]{ABG} to show
that $\K\subset\big(\dom(A),\H\big)_{1/2,1}$ and to get the
statement (d). The only non-trivial hypothesis one has to verify is
the inclusion $\dom(\<Q_3\>)\subset\dom(A)$, which has already been
shown in Lemma \ref{sur A}.
\end{proof}

Note that these results imply that $H$ has a spectral gap. We are
now ready to add a singular part to the perturbation $W$.

\section{Locally singular perturbations}\label{SectionSingular}
\setcounter{equation}{0}

In this section we deal with perturbations which are locally singular.
A particular attention is paid to Coulomb-type interactions. Our approach is deeply inspired from
\cite[Sec.~3]{Georgescu/Mantoiu}. In Lemma 3.4 of this reference,
the authors show that if $H$ and $\widetilde H$ are two selfadjoint
operators in $\H$ that coincide in some neighbourhood of infinity
and if one of them has a certain regularity property with respect to
the operator $Q$, then the difference of their resolvents is
short-range in the usual sense. This result is the key ingredient
for what follows.

Let us first recall some notations. If $\Lambda\subset\R^3$ is an
open set, then $H_\Lambda$ is defined as the restriction of the
selfadjoint operator $H$ to the subset
$\dom(H_\Lambda):=\{\psi\in\dom(H)\mid\supp(\psi)\subset\Lambda\}$.
We write $H_\Lambda\subset\widetilde H$ if for each
$\psi\in\dom(H_\Lambda)$ one has $\psi\in\dom(\widetilde H)$ and
$\widetilde H\psi = H\psi$. Next lemma is an application of the
abstract result mentioned above that takes \cite[Rem.~7.6.9]{ABG}
and the observation following \cite[Def. 2.16]{Georgescu/Mantoiu}
into account.

\begin{Lemma}\label{magique}
Let $H$ be as in Theorem \ref{thmborne}, and let $\widetilde H$ be a
selfadjoint operator in $\H$ such that $H_\Lambda \subset \widetilde
H$ for some neighbourhood $\Lambda\subset\R^3$ of infinity. Then for
each $z\in\C\setminus\{\sigma(H)\cup\sigma(\widetilde H)\}$ and for
each $\vartheta\in C^\infty\big([0,\infty)\big)$ with $\vartheta =0$
near $0$ and $\vartheta=1$ near infinity one has:
\begin{equation}\label{magnifique}
\int_1^\infty\d r\,\big\|\vartheta(|Q|/r)\big[(\widetilde
H-z)^{-1}-(H-z)^{-1}\big]\big\|<\infty.
\end{equation}
\end{Lemma}

\begin{proof}
Since the statement is an application of
\cite[Lem.~3.4]{Georgescu/Mantoiu} one only has to check its
non-trivial hypotheses, \ie (i) $\theta(Q)\dom(H)\subset\dom(H)$ for
all $\theta\in C^\infty_0(\R^3)$, and (ii) for all $\theta\in
C_0^\infty(\R^3\setminus\{0\})$ one has
$$
\int_1^\infty\d
r\left\{\left\|\[\theta(Q/r),H\]\right\|^2_{\dom(H)\to\H}
+\left\|\big[\theta(Q/r),[\theta(Q/r),H]\big]\right\|_{\dom(H)\to
\H}\right\}<\infty.
$$
Condition (i) follows from the identity
$$
\theta(Q)(H+i)^{-1}=(H+i)^{-1}\theta(Q)-i(H+i)^{-1}\alpha\cdot(\nabla\theta)(Q)(H+i)^{-1}
$$
valid on $\H$ (the proof of this relation is similar to that of
Lemma \ref{technique2} but simpler since $\theta$ is a bounded
function). For (ii) one observes that
$[\theta(Q/r),H]=ir^{-1}\alpha\cdot(\nabla\theta)(Q/r)$ and that
$\big[\theta(Q/r),[\theta(Q/r),H]\big] = 0$. Since
$\left\|\alpha\cdot(\nabla\theta)(Q/r)\right\|$ is bounded uniformly
in $r$ and since $r\mapsto r^{-1}$ belongs to $\ltwo([1,\infty),\d
r)$, one readily finishes the proof.
\end{proof}

Taking last lemma into account, we can prove that $H$ and
$\widetilde H$ have several similar properties.

\begin{Lemma}\label{yenamarre}
Let $H$ and $\widetilde H$ be as in Lemma \ref{magnifique}, and
assume that $\widetilde H$ is locally compact. Then $\sigma_{\rm
ess}(\widetilde H)=\sigma_{\rm ess}(H)$, $\widetilde H$ is of class
$C^{1,1}(A)$, the operator $A$ is strictly conjugate to
$\widetilde H$ on $[0,\infty)\setminus\{\sigma_{\rm
sym}^0\cup\sigma_{\rm pp}(H)\}$ and the operator $-A$ is strictly
conjugate to $\widetilde H$ on $(-\infty,0]\setminus\{\sigma_{\rm
sym}^0\cup\sigma_{\rm pp}(H)\}$.
\end{Lemma}

\begin{proof}
The difference $(\widetilde H+i)^{-1}-(H+i)^{-1}$ is a compact
operator due to \cite[Lem.~3.8]{Georgescu/Mantoiu} (the proof of
this result is based on the fact that both $H$ and $\widetilde H$
are locally compact and that $H$ has some regularity properties with
respect to the operator $Q$). This fact implies the first claim.

Since $H$ and $\widetilde H$ have the same essential spectrum and
$H$ has a spectral gap, these operators have a common spectral gap,
and thus there exists
$z\in\R\setminus\{\sigma(H)\cup\sigma(\widetilde H)\}$. Let
$R:=(H-z)^{-1}$ and $\widetilde R:=(\widetilde H-z)^{-1}$, then
$\widetilde R-R$ is compact. Furthermore, for each $\vartheta\in
C^\infty\big([0,\infty)\big)$ with $\vartheta=0$ near $0$ and
$\vartheta=1$ near infinity, it follows from Lemma \ref{magique}
that $\|\vartheta(|Q|/r)(\widetilde R-R)\|\in\lone([1,\infty),\d
r)$. Then an easy calculation shows that one also has
$$
\int_1^\infty\d r\,\big\|\vartheta(|Q_3|/r)(\widetilde
R-R)\big\|<\infty.
$$
By applying \cite[Thm.~7.5.8]{ABG} as in the proof of Proposition
\ref{V regularity}, it follows that $\widetilde R-R$ belongs to
$C^{1,1}(A)$. Now $R$ also belongs to $C^{1,1}(A)$ due to
Proposition \ref{V regularity}. Thus $\widetilde R$ belongs to
$C^{1,1}(A)$ and the second claim is proved.

The last claim is obtained from what precedes as in the proof of
Theorem \ref{thmborne}.
\end{proof}

Thus one only has to put into evidence non-trivial perturbations
$\widetilde H$ of $H$ such that the hypotheses of the previous
lemma are satisfied. For Coulomb perturbations of the free Dirac
operator without magnetic field, we recall that some care has to be
taken when choosing the selfadjoint extension to be considered (see
for example \cite{Arai/Yamada,Boutet/Purice,Klaus,Nenciu76} and
references therein). Such a difficulty also occurs when a magnetic
field is present. The treatment of this problem requires the
introduction of some notations. $\H^s:=\H^s(\R^3;\C^4)$, $s\in\R$,
is the usual Sobolev space of functions on $\R^3$ with values in
$\C^4$, $\E'(\R^3;\C^4)$ the set of compactly supported
distributions on $\R^3$ with values in $\C^4$, $\H^s_{\rm
c}(\R^3;\C^4):=\H^s\cap\E'(\R^3;\C^4)$, and $H_m$ is the free Dirac
operator $\alpha\cdot P+\beta m$ with domain $\H^1$ and form domain
$\H^{1/2}$. Finally, if $S$ is a selfadjoint operator in $\H$, we
recall that there exist a unitary operator $U$ and a positive
selfadjoint operator $|S|$ such that $S=U|S|=|S|U$. The form
associated with $S$ is then defined by
$$
h_S(\varphi,\psi):=\big\langle|S|^{1/2}\varphi,U|S|^{1/2}\psi\big\rangle,\qquad
\varphi,\psi\in\dom(h_S):=\dom\big(|S|^{1/2}\big).
$$
Next statement is a corollary of the main result of
\cite{Boutet/Purice} on selfadjoint extensions for the perturbed
Dirac operators. The behaviour of the potential at infinity is
prescribed by assumption (i) of Theorem \ref{MainThm}, and the local
regularity conditions of the potential are prescribed in assumption
(ii) of that theorem. In order to be consistent with the
notations of the introduction, we shall now write $H$ for the
``fully'' perturbed operator (which was previously denoted by
$\widetilde H$) and $H_{\rm reg}$ for the operator $H_0+ V_{\rm
reg}\equiv H_0+W$ (which was previously denoted by $H$ for
simplicity).

\begin{Proposition}\label{autoadjonction}
Assume that the hypotheses on $B$ and $V$ of Theorem \ref{MainThm}
hold. Then there exists a unique selfadjoint operator $H$ in $\H$,
formally equal to $H_0+V$, such that
\begin{enumerate}
\item[(a)] $\dom(H)\subset\H_{\rm loc}^{1/2}$,
\item[(b)] $\forall\varphi\in\dom(H)$ and $\psi\in\H_{\rm c}^{1/2}(\R^3;\C^4)$, one has
$\big\langle
H\varphi,\psi\big\rangle=h_{H_m}(\varphi,\psi)+h_{-\alpha\cdot
a+V}(\varphi,\psi)$.
\end{enumerate}
\end{Proposition}

\begin{proof}
In order to apply \cite[Thm.~1.3]{Boutet/Purice} one has to verify
the first two hypotheses of that theorem. The first one consists in
showing that for any $\phi\in C_0^\infty\big(\R^3,[0,1]\big)$ one
has $\phi(Q)(-\alpha\cdot a+V)\in\B(\H^{1/2},\H^{-1/2})$.
Fortunately, it is known that $\phi(Q)V_{\sf loc}$ is
$H_m$-bounded with relative bound $0$ and that $\phi(Q) V_{\sf c}$
is $H_m$-bounded with relative bound $2\nu$. Moreover $V_{\rm reg}$
belongs to $\B(\H)$ and the vector potential $a$ is in
$\linfloc(\R^2;\R^3)$. Thus the hypothesis $\phi(Q)(-\alpha\cdot
a+V)\in\B(\H^{1/2}, \H^{-1/2})$ is clearly fulfilled. It follows
that $H_m+\phi(Q)(-\alpha\cdot a+V)$ can be defined as an operator
sum in $\B(\H^{1/2},\H^{-1/2})$.

The second hypothesis requires that for any $\phi\in
C_0^\infty\big(\R^3,[0,1]\big)$ the operator
$H_\phi:=H_m+\phi(Q)(-\alpha\cdot a+V)$ defined on
$$
\dom_\phi:=\big\{\varphi\in\H^{1/2}\mid\big[H_m+\phi(Q)(-\alpha\cdot
a+V)\big]\varphi\in\H\big\}
$$
is a selfadjoint operator. Now, such a statement follows from
the main result of \cite{Nenciu76} and \cite{Nenciu77} (see also
\cite{Klaus}), which we recall in our setting: Under our assumptions
on $V$, there exists a unique selfadjoint operator $H^\phi$ such
that $\dom(H^\phi)\subset\H^{1/2}$ and
$$
\<H^\phi\varphi,\psi\>= h_{H_m}(\varphi,\psi)+h_{
\phi(Q)(-\alpha\cdot a+V)}(\varphi, \psi), \qquad\forall
\varphi\in\dom(H^\phi),\,\psi\in \H^{1/2}.
$$
Since $H_\phi$ has the same properties, then $H_\phi$ is equal to
$H^\phi$ by unicity, and the second hypothesis of
\cite[Thm.~1.3]{Boutet/Purice} is thus fulfilled.
\end{proof}

We can finally prove our main result.

\begin{proof}[Proof of Theorem \ref{MainThm}]
Clearly the operator $H_{\rm reg}= H_0 + V_{\rm reg}$ is selfadjoint
and satisfies the hypotheses of Theorem \ref{thmborne}. Let
$\Lambda\subset\R^3$ be a neighbourhood of infinity such that
$\Lambda\cap\supp(\chi)=\varnothing$. Then, using the definitions of
$\dom[(H_{\rm reg})_\Lambda]$, $\dom(H)$ and $V_{\rm reg}$, we get
\begin{align*}
\dom[(H_{\rm reg})_\Lambda]&=\big\{\varphi\in\H^{1/2}_{\rm loc}\mid
H_m\varphi
+(-\alpha\cdot a+V_{\rm reg})\varphi\in \H,~\supp(\varphi)\subset\Lambda\big\}\\
&=\big\{\varphi\in\H^{1/2}_{\rm loc}\mid H_m\varphi
+(-\alpha\cdot a+V)\varphi\in\H,~\supp(\varphi)\subset\Lambda\big\}\\
&\subset\big\{\varphi\in \H^{1/2}_{\rm loc}\mid H_m\varphi
+(-\alpha\cdot a+V)\varphi\in\H\big\}\\
&=\dom(H).
\end{align*}
Thus, the property $(H_{\rm reg})_\Lambda\subset H$ holds.
Furthermore the operator $H$ is locally compact due to the inclusion
$\dom(H)\subset\H^{1/2}_{\rm loc}$  (Proposition
\ref{autoadjonction}.(a)). Thus the couple $(H_{\rm reg},H)$
satisfies both hypotheses of Lemma \ref{yenamarre}. Then, statement
(a) follows from Lemma \ref{compactness} and from the first
assertion of Lemma \ref{yenamarre}. Statements (b) and (c) follow
from the other assertions of Lemma \ref{yenamarre} and from the
abstract conjugate operator method
\cite[Cor.~7.2.11\,\&\,Thm.~7.4.2]{ABG}. Statement (d) is obtained
as in Theorem \ref{thmborne}.
\end{proof}

\section*{Acknowledgements}

The idea of Lemma \ref{rhoegalite} is due to V. Georgescu. We
express our deep gratitude to him for this and for many other
precious discussions. S. R. thanks B. Helffer for a two weeks
invitation to Orsay where part of the present work was performed.
This stay was made possible thanks to the European Research Network
``Postdoctoral Training Program in Mathematical Analysis of Large
Quantum Systems" with contract number HPRN-CT-2002-00277. R. T. d.
A. thanks the Swiss National Science Foundation for financial
support.

\section*{Appendix}

\begin{proof}[Proof of Lemma \ref{technique1}]
(a) Let $\varphi$, $\psi$ be in $\D$. Using the transformation
\eqref{F unitaire}, one gets
$$
\<H_0^{-n}\varphi,Q_3\psi\>=\int_\R\d\xi\<H_0(\xi)^{-n}(\F\varphi)(\xi),
(i\partial_\xi\F\psi)(\xi)\>_{\H_{\scriptscriptstyle 12}}.
$$
Now the map $\R\owns\xi\mapsto
H_0(\xi)^{-n}\in\B(\H_{\scriptscriptstyle 12})$ is norm
differentiable, with its derivative given by
$$
-\sum_{j=1}^nH_0(\xi)^{-j}\alpha_3H_0(\xi)^{j-n-1}.
$$
Hence the collection
$\{\partial_\xi[H_0(\xi)^{-n}(\F\varphi)(\xi)]\}_{\xi\in\R}$ belongs
to $\int_\R^\oplus\d\xi\,\H_{\scriptscriptstyle 12}$. Thus one can
perform an integration by parts (with vanishing boundary
contributions) and obtain
$$
\<H_0^{-n}\varphi,Q_3\psi\>=\int_\R\d\xi\<i\partial_\xi[H_0(\xi)^{-n}(\F\varphi)(\xi)],
(\F\psi)(\xi)\>_{\H_{\scriptscriptstyle 12}}.
$$
It follows that
$\left|\<H_0^{-n}\varphi,Q_3\psi\>\right|\le\textrm{Const.}\:\!\|\psi\|$
for all $\psi\in\D$. Since $Q_3$ is essentially selfadjoint on $\D$,
this implies that $H_0^{-n}\varphi$ belongs to $\dom(Q_3)$. The
second statement can be proved using a similar argument.

(b) The boundedness of $P_3|H_0|^{-1}$ is a consequence of the
estimate
$$
\esssup_{\xi\in\R}\|\xi |H_0(\xi)|^{-1}\|_{\B(\H_{\scriptscriptstyle
12})}
=\esssup_{\xi\in\R}\big\|\xi[H_0(0)^2+\xi^2]^{-1/2}\big\|_{\B(\H_{\scriptscriptstyle
12})} <\infty
$$
and of the direct integral formalism. The remaining assertions
follow by standard arguments.
\end{proof}

\begin{proof}[Proof of Lemma \ref{technique2}]
The first statement is easily obtained by using the equality
$g(x)=g(0)+\int_0^x\d y\,g'(y)$. For the second one, let us observe
that the following equality holds on $\D$:
\begin{equation}\label{Naudts}
H_0^{-1}g(Q_3)H_0=g(Q_3)+iH_0^{-1}\alpha_3g'(Q_3).
\end{equation}
Now, for $\varphi,\psi\in\D$ and $\eta\in H_0^{-n}\D$, one has
\begin{align*}
&\<\varphi,H_0^{-1}g(Q_3)\eta\>-\<\varphi,g(Q_3)H_0^{-1}\eta\>\\
&=\<\varphi,H_0^{-1}g(Q_3)H_0\psi\>+\<\varphi,H_0^{-1}g(Q_3)(\eta-H_0\psi)\>
-\<\overline g(Q_3)\varphi,H_0^{-1}\eta\>\\
&=\<\varphi,g(Q_3)\psi\>+\<\varphi,iH_0^{-1}\alpha_3g'(Q_3)\psi\>
+\<\overline g(Q_3)H_0^{-1}\varphi,(\eta-H_0\psi)\>\\
&\qquad-\<\overline g(Q_3)\varphi,H_0^{-1}\eta\>\\
&=\<\overline g(Q_3)\varphi,H_0^{-1}(H_0\psi-\eta)\>
+\<\varphi,iH_0^{-1}\alpha_3g'(Q_3)H_0^{-1}\eta\>\\
&\qquad+\<\varphi,iH_0^{-1}\alpha_3g'(Q_3)H_0^{-1}(H_0\psi-\eta)\>
+\<\overline g(Q_3)H_0^{-1}\varphi,(\eta-H_0\psi)\>,
\end{align*}
where we have used \eqref{Naudts} in the second equality. Hence
there exists a constant $\textsc c>0$ (depending on $\varphi$) such
that
$$
\big|\<\varphi,H_0^{-1}g(Q_3)\eta\>-\<\varphi,g(Q_3)H_0^{-1}\eta\>
-\<\varphi,iH_0^{-1}\alpha_3g'(Q_3)H_0^{-1}\eta\>\big|\le\textsc
c\:\!\|\eta-H_0\psi\|.
$$
Then the statement is a direct consequence of the density of $H_0\D$
and $\D$ in $\H$.
\end{proof}

\begin{proof}[Proof of Lemma \ref{technique3}]
The invariance of the domain of $Q_3$ follows from the fact that
$f(P_3)\in C^1(Q_3)$. Thus the expression
$f(P_3)g(Q_3)-g(Q_3)f(P_3)$ is well-defined on $\dom(Q_3)$.
Moreover, by using the commutator expansions given in
\cite[Thm.~5.5.3]{ABG}, one gets the following equality in form
sense on $\D$:
\begin{equation}\label{bientotlabouffe}
f(P_3)g(Q_3)-g(Q_3)f(P_3)=-i\int_0^1\d\tau\int_\R\d x\,\e^{iP_3\tau
x}g'(Q_3)\e^{iP_3(1-\tau)x} \widehat{f'}(x).
\end{equation}
Since the r.h.s. extends to a bounded operator, and since
$\D\subset\dom(Q_3)$ is a core for $g(Q_3)$, the second statement
follows.

The last statement is obtained by proving that the operator
$\<Q_3\>^s[f(P_3),g(Q_3)]\<Q_3\>^{-s}$, defined in form sense on
$\D$, extends to a bounded operator. Again, by using the explicit
formula \eqref{bientotlabouffe}, the submultiplicative property of
the function $\<\:\!\cdot\:\!\>$ and the hypothesis on the map
$x\mapsto\<x\>^s\widehat{f'}(x)$, this result is easily obtained.
\end{proof}



\end{document}